# Model of a transition from low to high ablation regime in a carbon arc

A. Khrabry[1,*], I. D. Kaganovich[1], A. Khodak[1], V. Vekselman[1,†], T. Huang[1]

[1]*Princeton Plasma Physics Laboratory, Princeton NJ 08542*

## Abstract

Graphite ablation in a presence of inert background gas is widely used in different methods for the synthesis of carbon nanotubes, including electric arc and laser/solar ablation. The ablation rate is an important characteristic of the synthesis process. It is known from multiple arc experiments that there are two distinguishable ablation regimes, so-called "low ablation" and "high ablation" regimes in which the ablation rate behaves rather differently with variation of the arc parameters. We developed a model that explains low and high ablation regimes by taking into account the presence of a background gas and its effects on the ablation rate. We derive analytical relations for these regimes and verify them by comparing them with full numerical solutions in a wide arc parameter range. We comprehensively validate the model by comparing to multiple experimental data on the ablation rate in carbon arcs, where various arc parameters were varied. Good qualitative and quantitative agreement between full numerical solutions, analytical solutions, and experimental data was obtained.

## I. Introduction

Gas-phase production of carbon nanoparticles is often based on evaporation of bulk solid graphite into an atmosphere of an inert background gas where evaporated carbon condenses and serve as a feedstock for the nanoparticle growth[1,2,3,4,5,6,7,8,9,10,11]. Electric arc[1,2,3,4], or a laser/focused solar light beam[6,7,8,12] is utilized as a heat source for the ablation. In particular, the electric arc with graphite electrodes (i.e. carbon arc) is known as a cheap and scalable method to produce high-quality carbon nanotubes[13,14,15]. The ablation rate of a graphite anode is an important characteristic of the synthesis process that determines the synthesis yield and affects the gas phase conditions for the growth of the nanoparticles such as gas composition, temperature, and residence time.

Multiple experiments have been conducted with carbon arcs for nano-synthesis[4,16,17,18]. In these experiments, arcs were run between two coaxial cylindrical electrodes in an atmosphere of helium at 500 torr. Radii of electrodes and their separation were varied as well as the arc current. Experiments showed that the ablation rate of the graphite anode is a complex function of the arc parameters; two distinguishable regimes were observed: high and low ablation regimes. As an example, experimental

---

data[18] on the ablation rate as a function of arc current at fixed anode size and electrode separation is plotted in Fig. 1. At low current, the ablation rate is low and weakly depends on the arc parameters (this is a low ablation regime). However, when the arc current exceeds certain value, the ablation rate increases rapidly with the arc current (this is a high ablation regime). Similar behavior of the ablation rate, with low ablation and high ablation regimes, was also observed for variation of the anode radius and the inter-electrode gap. A model capable to explain both ablation regimes and predict the ablation rate as a function of experimental parameters for an electric arc (or a laser or UV light beam) is needed for planning the experiments and to better control the synthesis process.

Multiple computational and theoretical models have been developed for carbon arcs[2,17,19,20,21]. However, these models had a focus rather on the arc plasma than on the ablation rate: ablation rate was either pre-defined or was estimated using Langmuir's law[22] for evaporation in vacuum. As a result, these models cannot describe and predict both high and low ablation regimes with a rapid transition between them. It is known that the presence of background gas can substantially limit the outflow of the evaporated material from the evaporation surface[9,23], thereby reducing the ablation rate. Our latest model[24] has a more accurate expression for the ablation rate that accounts for the presence of the background gas, however, the results were obtained for the low ablation regime only.

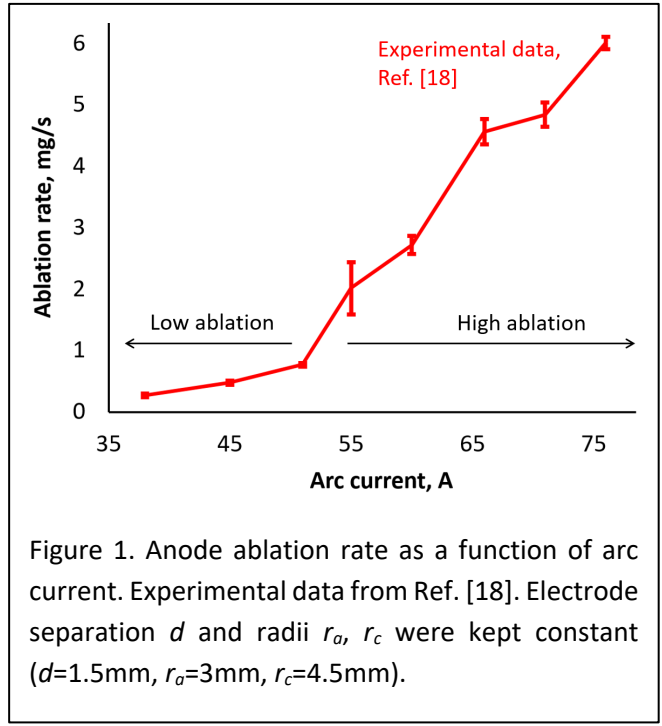

Figure 1. Anode ablation rate as a function of arc current. Experimental data from Ref. [18]. Electrode separation $d$ and radii $r_a$, $r_c$ were kept constant ($d$=1.5mm, $r_a$=3mm, $r_c$=4.5mm).

In this paper, we theoretically explain that the presence of a background gas plays a crucial role in the formation of two ablation regimes observed in the experiments. We derive an ablation model based on heat balance in between a plasma and an electrode (or a target for the light beam) and provide analytical expressions for the ablation rate of a thin cylindrical electrode.

The paper is organized as follows. In section II, the ablation model accounting for the presence of a background gas is described, an explanation for the low and high ablation regimes is given, analytical expressions for the ablation flux and the surface temperature corresponding to the transition between two ablation regimes are derived. Section III provides validation of the ablation model by comparison to the available experimental data on the ablation rate as a function of the surface temperature. In Section IV, a model to determine anode ablation rate from the energy balance in a carbon arc is described. Section V presents analytical expressions for ablation rate in high and low ablation regimes which are derived as functions of the arc parameters. In section VI, numerical and analytical solutions are validated via a comparison to vast experimental data on the anode ablation rate in carbon arcs with various arc currents, electrode separation distances and anode radii. Section VII contains the summary of the work.

## II.     The ablation model

### II.a. Ablation in the presence of a background gas

In case of ablation in a vacuum, a flux of ablated material from a surface is given by Langmuir's formula[22]:

$$g_{abl}(T_a) = p_{sat,C}(T_a)/v_{th}, \quad v_{th} = \sqrt{2\pi k T_a/m_C}. \tag{1}$$

Here, $m_C$ is the mass of ablated particles (carbon atoms or molecules), $k$ is the Boltzmann constant, $T_a$ is the surface temperature, and $p_{sat,C}(T_a)$ is the saturation pressure of carbon given by the Clausius-Clapeyron relation:

$$p_{sat,C}(T_a) = p_0 \exp\left(-\frac{Lm_C}{kT_a}\right), \tag{2}$$

where $L$ is the latent heat of graphite ablation, $p_0$ is a material-dependent constant. Values of the ablation flux parameters $L$, $m_C$ and $p_0$ are discussed in Appendix I and are also summarized in Appendix III.

If there is a background gas, it will impede the flow of the ablated material away from the ablating surface [9,23] leading to its accumulation near the surface. This will create a returning flux of the ablated material back to the evaporation surface reducing the net ablation flux[24,25]:

$$g_{abl} = (p_{sat,C}(T_a) - p_{C,a})/v_{th}. \tag{3}$$

Here, $p_{C,a}$ is the partial pressure of carbon vapor at the ablating surface. The pressure of carbon vapor is determined by the diffusion of ablated material through the background gas. As evident from Eq. (3), the background gas can notably affect the net ablation flux via a build-up of the near-surface carbon pressure $p_{C,a}$ which is determined by the diffusion of the carbon vapor through the background gas and the ablation flux, $g_{abl}$.

Typically, in nano-synthesis arcs or light-beam (laser) ablation systems, condensation of the ablated carbon takes place rather close to the ablating surface (see Fig. 2(b)). In electric arcs most of the ablated material deposits at a closely located cathode[18,24]. In laser-ablation systems, the temperature rapidly decays below the saturation point $T_{sat} = -Lm_C/[k \ln(p/p_0)]$ (on a scale of about 1mm[8]) leading to rapid carbon vapor condensation. In other words, the diffusion length is smaller than the ablation surface width. This allows us to employ a 1D formulation for the diffusion of carbon. A constant carbon flux, $g_{abl}$, is driven through stationary helium by a gradient of the carbon pressure, as described by the Stefan-Maxwell equation[26,27,28,29] (see Appendix II for the details):

$$\frac{dp_C}{dx} = -\frac{p_{He}}{nD_{C-He}}\frac{g_{abl}}{m_{C,a}}. \tag{4}$$

Here, $p_C$ and $p_{He}$ are the carbon and helium partial pressures, $n = p/kT$ is the gas mixture density, and $D_{C-He}$ is the binary diffusion coefficient, which can be obtained using the kinetic theory of gases[26,27,28,29], and $m_{C,a}$ is the mass of carbon atoms. Carbon molecules are only present in very thin regions near the arc electrodes[42] and, therefore, don't play important role in the carbon transport in the arc volume.

Thereby, only carbon atoms are considered in Eq. (4). Note that the denominator $nD_{C-He}$, as defined by (AII.2), is a weak function of temperature. For typical temperature in an electric arc, $nD_{C-He} = 5 \cdot 10^{21}\ m^{-1}s^{-1}$.

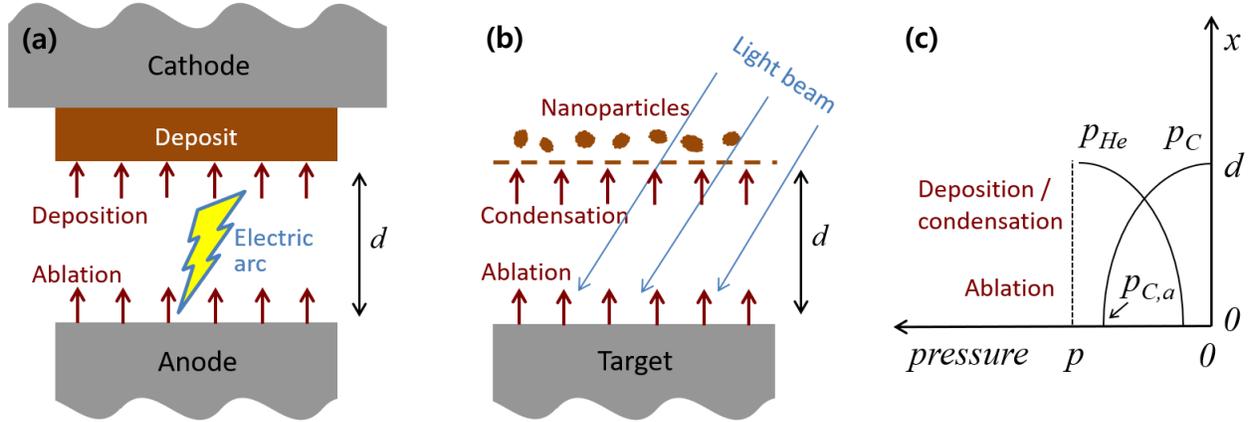

*Figure 2. Schematic of carbon transport in an electric arc (a) and a light ablation system (b); schematic profiles of carbon and helium partial pressures (c). Near the ablation surface of an electric arc anode or an irradiated target, carbon pressure is at its maximum: $p_C = p_{C,a}$. In the condensation/deposition region (at the cathode of electric arc or in the nanoparticles formation region in case of a light beam ablation), carbon pressure is close to zero. The carbon pressure gradient drives its diffusion through the background gas, i.e. helium. The total pressure in the system is equal to a sum of carbon and helium partial pressures and is constant: $p_C + p_{He} = p = const$.*

As discussed in Appendix II, at conditions typical for the arc and light-beam ablation experiments, variation of the total pressure $p$ is small, and the fraction of electrons and ions in the gas mixture (the ionization degree) is low. Thereby, the sum of carbon and helium partial pressures $p_C + p_{He} \approx p$ can be taken constant, and Eq. (4) can be solved for $p_C$. Zero carbon density can be specified as a boundary condition at the surface opposite to the ablating wall (either a cathode surface in an electric arc or a condensation layer in the laser/solar ablation case, see Fig. 2). Solution of Eq. (4) yields following carbon gas pressure at the ablating surface:

$$p_{C,a} = p\left[1 - \exp\left(-\frac{g_{abl}}{g_0}\right)\right]. \qquad (5)$$

Here, $g_0$ is a characteristic carbon flux determined by the diffusion:

$$g_0 = \frac{m_{C,a} nD_{C-He}}{d}. \qquad (6)$$

Here, $d$ is the inter-electrode gap size (or a distance to the condensation region in case of laser/solar ablation; see Fig. 2). Substitution of (5) into (3) yields following expression with only one unknown, the ablation flux $g_{abl}$:

$$g_{abl} = \frac{p_{sat,C}(T_a)}{v_{th}(T_a)} - \frac{p}{v_{th}(T_a)}\left[1 - exp\left(-\frac{g_{abl}}{g_0}\right)\right]. \qquad (7)$$

As predicted by Eq. (7), $g_{abl}$ depends on the surface temperature $T_a$, background pressure $p$, condensation distance $d$ and material properties. Eq. (7) can be solved for $g_{abl}$ numerically. The solution of Eq. (7) and the corresponding carbon pressure $p_{C,a}$ are plotted in Fig. 3 with green lines as functions of surface temperature $T_a$ for the background pressure $p$ = 500 torr and two inter-electrode gap widths d = 1 mm and d = 5 mm (which cover the range in which the gap was varied in most experiments) corresponding to $g_0$ = 0.1 kg/m²/s and 0.02 kg/m²/s, respectively.

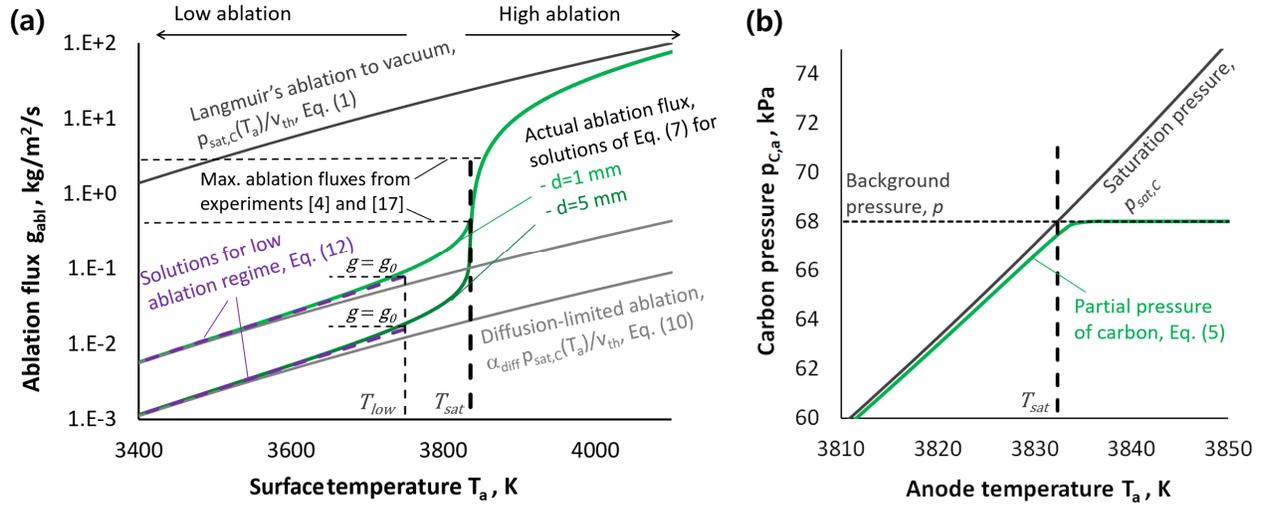

Figure 3. Ablation flux $g_{abl}$ (a) and partial carbon pressure at the surface $p_{C,a}$ (b) as functions of surface temperature $T_a$ for background pressure p=500 torr and inter-electrode gap widths d=1mm and d=5mm. Exact numerical solution of Eqs. (5) and (7) are shown by green lines, asymptotic behavior is shown by grey lines. At low temperatures, below the temperature $T_{low}$ given by Eq. (11), the ablation rate is substantially reduced by the presence of background gas (see Eq. (9)). The reason for this is a build-up of carbon pressure $p_{C,a}$ which is close to the saturation pressure; a small difference determines the net ablation flux. At the saturation temperature $T_{sat} = -Lm_C/[k\ln(p/p_0)]$, partial carbon pressure $p_{C,a}$ reaches its limit determined by the background pressure, and can no longer suppress the ablation. The ablation flux grows drastically. At higher temperatures, the role of background gas is small; ablation flux approaches Langmuir's solution, Eq. (1).

As evident from the figure, two regimes corresponding to high and low ablation rates can be distinguished with a rapid transition between them. The underlying physics of these regimes is discussed below, analytical expressions for the surface temperature, ablation flux and the transition temperature are derived.

Using the fact that $p_{sat,C}$ and $v_{th}$ are, respectively, a strong and a weak function of $T_a$, $T_a$ can be explicitly expressed from Eq. (7) as a function of the ablation flux $g_{abl}$ and the background pressure $p$ by substituting a reference value for $T_a = T_{sat}$ in $v_{th}$:

$$\frac{1}{T_a} = \frac{1}{T_{sat}} - \frac{k}{Lm_C} \ln\left(\frac{g_{abl}}{p}\sqrt{\frac{2\pi k T_{sat}}{m_C}} + \left[1 - exp\left(-\frac{g_{abl}}{g_0}\right)\right]\right). \tag{8}$$

Here, $T_{sat} = -Lm_C/[k\ ln(p/p_0)]$ is the saturation temperature. For the temperature range plotted in Fig. 3(a), an error of $T_a$ given by relation (8) is very small, below 4 K.

### II.b. Low ablation regime

At surface temperatures lower than the saturation temperature of carbon vapor, $T_a < T_{sat}$, and, correspondingly, the carbon vapor saturation pressure lower than the system pressure, $p_{sat,C}(T_a) \ll p$, ablation flux is lower than the characteristic diffusive flux, $g_{abl} \ll g_0$. With this, the left-hand side of Eq. (7) can be neglected and exponent in the right-hand side (RHS) can be linearized yielding

$$g_{abl} \approx g_0 \frac{p_{sat,C}(T_a)}{p}, \tag{9}$$

or, using definition of $g_0$ given by Eq. (6):

$$g_{abl}(T_a) \approx \alpha_{diff} \frac{p_{sat,C}(T_a)}{v_{th}} = \frac{p_{sat,C}(T_a)}{p} m_{C,a} n_0 v_{th}. \tag{10}$$

Here,

$$\alpha_{diff} = \frac{3\pi}{8}\sqrt{\frac{m_{C,a}+m_{He}}{2m_{He}}}\frac{kT}{p\sigma_{C-He}d} = 1.7\frac{kT}{p\sigma_{C-He}d}, \text{ and}$$

$$n_0 = \frac{3}{16}\sqrt{2\frac{m_{C,a}+m_{He}}{m_{He}}}\frac{1}{\sigma_{C-j}d} = \frac{0.53}{\sigma_{C-j}d}.$$

is the coefficient of ablation flux reduction due to the diffusion of ablated products through the background gas, i.e. helium; $n = p/kT$ is the gas mixture density.

As evident from Eq. (10) and Fig. 3(a), ablation flux is drastically reduced by the background gas at low temperatures. Ablation flux is proportional to one that would be in case of ablating into vacuum (Eq. (1)) with a very small factor $\alpha_{diff}$. For typical arc parameters, d=2 mm and p=500 torr, $\alpha_{diff}$ is about 0.003. The larger is the electrode separation $d$ and the pressure p, the smaller are $\alpha_{diff}$ and the ablation flux.

Figure 3(b) shows the carbon pressure at the surface, $p_{C,a}$ as a function of the surface temperature. As evident form the figure, at $T_a \approx T_{sat}$, the carbon pressure at the surface, $p_{C,a}$ is close to the saturation pressure $p_{sat,C}(T_a)$ for $T_a \leq T_{sat}$. This result can be formally obtained by substituting (9) into (5) with a linearized RHS. In this regime, carbon evaporation is almost completely balanced by the redeposition of ablated material. As a result, the net ablation flux is very low as predicted by Eq. (10).

These conditions correspond to the low ablation regime observed in the arc experiments. As evident from Fig 3(a), approximations (9) and (10) work reasonably well when the ablation flux is lower than $g_0$.

According to Eq. (7) with the LHS neglected, ablation flux $g_{abl} = g_0$ corresponds to the surface temperature $T_a = T_{low}$ equal to:

$$T_{low} = \frac{T_{sat}}{1+ln[1-exp(-1)] / ln(p/p_0)}, \qquad (11)$$

where $T_{sat} = -Lm_C/[k\, ln(p/p_0)]$ is the saturation temperature. Values of $T_{sat}$ and $T_{low}$ for the background pressure used in the arc experiments are given in Appendix III.

The lower is the surface temperature $T_a$ the more accurate is the approximation (9) which was derived from Eq. (7) by linearization of its RHS. For $T_a = T_{low}$ corresponding to $g_{abl} = g_0$, from Eq. (7) it follows that $p_{sat,C}/p = 1 - 1/e$, and an error of Eqs. (9) and (10) is about 40%.

More accurate expression for the ablation rate can be derived using the second order Taylor expansion of the exponent in RHS of Eq. (7):

$$g_{abl} = g_0 \left( \frac{p_{sat,C}(T_a)}{p} + \frac{1}{2}\left(\frac{p_{sat,C}(T_a)}{p}\right)^2 \right). \qquad (12)$$

The results of Eq. (11) are shown in Fig. 3(a) with dashed lines. This approximation works well for surface temperatures up to $T_{low}$.

### II.c. High ablation regime

Partial pressure of carbon at the surface $p_{C,a}$, apparently, cannot be higher than the background pressure $p$ in the system. When the surface temperature becomes higher than one of carbon saturation, $T_{sat}$, the partial pressure of carbon $p_{C,a}$ reaches its limit $p$ and the deposition flux from the gas can no longer compensate for the ablation flux from the surface. At this point, the net ablation flux grows drastically with the surface temperature, by more than an order of magnitude with the temperature increase of only 10 K, as denoted with a vertical dashed line in Fig. 3(a). This behavior explains the rapid growth of the oblation rate in the high ablation regime, as observed in the experiments.

At even higher temperatures, $T_a > T_{sat}$, corresponding to $p_{sat,C}(T_a) \gg p$, the carbon pressure at the surface $p_{C,a}$ becomes negligible as compared to $p_{sat,C}$. The ablation flux approaches one given by Langmuir's formula (1) for evaporation in vacuum. For the background pressure of 500 torr, this regime corresponds to the ablation flux above $10\ kg/(m^2s)$ (see Fig. 3(a)) and flow velocity $v = g_{abl}kT/p_C m_C$ above 400 m/s. At such velocities, the total pressure $p$ must be adjusted to account for fast gas flow with high Mach numbers; i.e., the constant pressure approximation, $p = const$, cannot be used. However, the carbon pressure will still be negligible as compared to $p_{sat,C}$ in this case. Such an ablation flux corresponds to the ablation rate of 300 mg/s from a 3 mm radius surface (typical radius of an electrode in arc experiments). This regime was not reached in any known arc or laser/solar ablation experiment. For comparison, in experiments from Refs. [4,8,17,18,30] maximum ablation rates were below 50 mg/s. The highest ablation rate of 50 mg/s was observed in Ref. [30] with a powder-filled anode, which is not considered here. Maximum ablation fluxes observed in experiments[4,17] is shown for comparison in Fig. 3(a) with horizontal dashed lines.

## III. Validation of the ablation flux relation (10)

Experimental data on the ablation rate as a function of THE surface temperature is available in Ref. [8]. The front surface of a cylindrical graphite target of a radius of 3 mm was heated up by a continuous wave laser in a helium atmosphere of 230 torr. Various admixtures to the graphite were tried in Ref. [8] with virtually no effect on the ablation rate. The measurements were performed for temperatures up to 3600 K, which is lower than $T_{low}$ (3630 K for 230 torr). Correspondingly, we should expect a low ablation regime in which the ablation rate is much lower than one predicted by Langmuir's ablation formula (1), with a constant factor difference $\alpha_{diff}$, as given by Eq. (10). This is exactly what we see in Figure 4, with a convenient choice of axis (natural logarithm of ablation rate and inversed temperature, similar to Ref. [8]) that converts relations (1) and (10) into straight lines. Experimental points are lower than Langmuir's curve by a constant factor of $\alpha_{diff} = 0.01$.

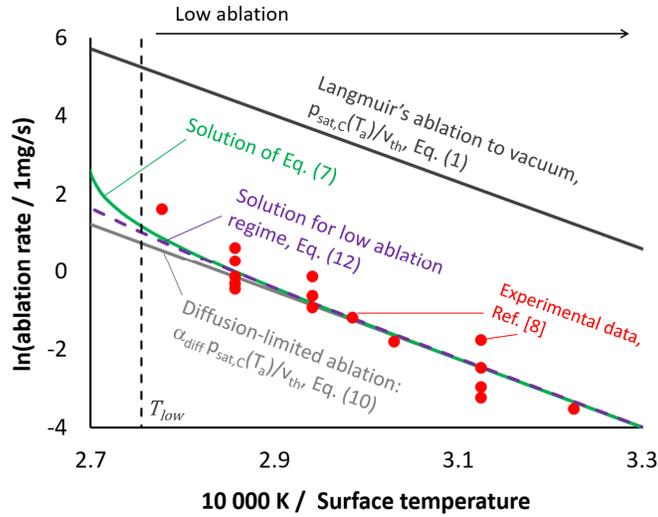

Figure 4. Ablation rate as a function of surface temperature. Comparison between experimental data[8], Langmuir's formula (1) for evaporation in vacuum and solutions (7), (10), and (12) accounting for the ablation reduction by the background gas. Approximate solutions for low ablation regime (10), (12) are close to the full solution (6) and are in a good agreement with the experimental data.

A reduction factor $\alpha_{diff}$ of 0.01 corresponds to d=1mm, distance from the target front surface to a location where the carbon vapor condenses (see Fig. 2). This value agrees with the nanoparticles detection results from: 1mm is the closest distance to the ablating surface where carbon nanoparticles were detected indicating rapid condensation of carbon. This distance is also in a good agreement with a temperature profile presented in Ref. [8] showing that at a distance 1mm from the surface, temperature is reduced by 300 K which typically corresponds to rapid condensation of a vapor[31].

# IV. Determining anode temperature and ablation rate from the energy balance

The ablation rate can be determined from the energy balance at the ablating surface in a carbon arc, i.e. the anode front surface, see Fig. 5. Sum of the heat sources at the surface must be equal to the sum of heat sinks. List of the heat sinks and sources are provided below. Heat transfer through the anode body is considered in one-dimensional approximation valid for long cylindrical anodes typically used in arc experiments. Values of the model parameters are summarized in Appendix III. Additionally, in this section, remarks will be made on how this consideration can be applied to laser/solar ablation systems.

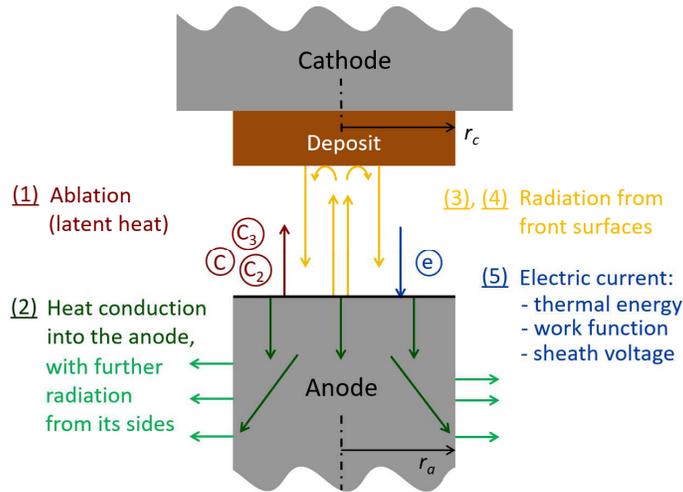

*Figure 5. Energy balance schematic at the anode front surface in carbon arc. Equalizing the sum of heat sources at the surface to the sum of heat sinks yields the energy balance equation (13).*

Heat sinks at the anode front surface are:

(1) The heat sink corresponding to the ablation of the anode material (main object of the paper interest) is

$$L \cdot \pi r_a^2 g_{abl},$$

where $r_a$ is the radius of the cylindrical anode; $L$ is the latent heat of graphite evaporation.

(2) The heat conduction into the anode body along the electrode's axis with further radiation from its side surface is given by:

$$\pi T_a^{2.5} r_a^{1.5} \sqrt{4/5 \cdot \sigma \varepsilon_a \lambda_a},$$

where $T_a$ is the temperature of the anode front surface, $\sigma$ is Stefan-Boltzmann radiation constant, $\lambda_a$ is thermal conductivity of the anode (assumed constant along the electrode), $\varepsilon_a$ is the emissivity of the anode surface.

This analytical expression for heat transfer in a long cylinder was obtained in Ref. [32], where it was also shown that the heat conduction to ambient gas and Joule heating in the electrode play a negligible role in determining heat flux through the anode front surface.

(3) The radiation energy loss from the front surface is given by:

$$\pi r_a^2 \sigma \varepsilon_a T_a^4 (1 - \alpha_r).$$

Here, $\alpha_r = (1 - \varepsilon_c)\varepsilon_a F_{a \to c}^2$ is a small coefficient accounting for the radiation reflected from the cathode front surface back to the anode (subsequent reflections between the anode and the cathode are neglected). $\varepsilon_c$ is the emissivity of the cathode surface. $F_{a \to c}$ is a geometrical view factor between the cathode and anode front surfaces representing a fraction of radiation energy emitted from the anode front surface that reaches the cathode. View factor between two coaxial discs is given by[17,33]:

$$F_{a \to c} = \left(x - \sqrt{x^2 - 4 r_a^2/r_c^2}\right)/2,$$

$$x = 1 + d^2/r_c^2 + r_a^2/r_c^2.$$

Here, $r_c$ is radius of the cathode deposit that covers the cathode with a thick layer during the arc run and forms new front surface reflecting and emitting radiation. This surface is also circular but narrower than the cathode. Experiments have shown that if a cathode is much wider than the anode, as it was in Refs. [16,17], with cathodes up to several centimeters wide, then the deposit is also considerably wide (see figures in these references) and view factor $F_{a \to c}$ is close to unity. For relatively narrow cathodes used in Refs. [1,3,18], (diameter 9mm), the cathode deposit is typically of the anode width ($r_c = r_a$). With a typical electrode separation $d = 1.5 mm$, the view factor is $F_{a \to c} = 0.6$.

Terms (1) – (3) are similar in the cases of electric arc and laser/solar ablation of a cylindrical target (as long as a target has a cylindrical shape, as in Refs. [7, 8]); $\alpha_r$ is zero in the light ablation cases.

The heat sources at the anode front surface are:

(4) The thermal radiation from the cathode front surface reaching the anode. Making use of a relation for mutual view factors ratio[33], $F_{c \to a} = F_{a \to c} r_a^2/r_c^2$, this heat source term can be written as:

$$\pi r_a^2 \sigma \varepsilon_a (\varepsilon_c F_{a \to c} T_c^4).$$

Here, $T_C$ is the temperature of the cathode front surface equal to 3400 K to support thermionic emission[24,29].

(5) Electrons bring thermal energy from the plasma, energy gained in the space-charge sheath and work function[20,29,34,35,36]:

$$\left(V_w + 2.5 \frac{k}{e} T_{e,a} + \max(V_{sh}, 0)\right) I.$$

Here, $V_w$ is the work function of the anode material (4.6 eV for graphite), $T_{e,a}$ is the electron temperature in plasma at the anode surface, $V_{sh}$ is the voltage drop within the anode space-charge sheath and $I$ is the arc current. The ion current at the anode surface is negligible[32], hence, the electron current at the anode surface can be taken equal to the arc current.

As shown in Ref. [21] with 1D modeling of carbon arc accounting for ablation and chemical transformations in the gas phase, electron temperature near the anode surface $T_{e,a}$ is about 1.7 eV, very weakly dependent on the ablation rate and other arc parameters. The value is typically lower for arcs with no anode ablation. Ablated carbon has to be ionized; carbon molecules need to be dissociated first. These processes require a lot of energy implying a high electric field near the anode and high electron temperature.

The anode sheath voltage in atmospheric pressure arcs is usually negative, of about 1 or 2 Volts[20,37,38,39]. However, when a surface of an electrode is hot, the electron emission can substantially change the sheath voltage and even inverse its sign[40,41]. As shown in Refs. [29,32], if the anode is hot (which is the case when graphite ablates), the anode sheath voltage $V_{sh}$ is positive to suppress the electron emission from its surface. The anode sheath voltage is proportional to the anode temperature, and is roughly $0.5\ V$ corresponding to the temperature of graphite ablation, $T_{sat}$.

With $V_w$, $T_{e,a}$ and $V_{sh}$ being almost constant, the electron heat flux expression can be simplified to:

$V_{eff} \cdot I$,

with the effective voltage $V_{eff} = 9.5\ V$.

The gas temperature gradient near the anode is small[21, 29,32], and the thermal conduction through the gas can be neglected.

For laser/solar ablation, terms (4) and (5) are irrelevant. In these cases, the heat source takes a simple form, $\varepsilon_a \cdot Q_{light}$, where $Q_{light}$ is the energy of the light beam to the target from the surface.

Resultant energy balance equation:

Equalizing the sum of heat sources at the front surface to the sum of heat sinks yields following the energy balance equation:

$$L \cdot 4\pi r_a^2 g_{abl}(T_a) + C_1 T_a^{2.5} r_a^{1.5} + C_2 r_a^2 T_a^4 = V_{eff} \cdot I + C_3 r_a^2. \tag{13}$$

Here, constants $C_1 = \pi\sqrt{4/5 \cdot \sigma \varepsilon_a \lambda_a}$, $C_2 = \pi \sigma \varepsilon_a (1 - \alpha_r)$ and $C_3 = \pi \sigma \varepsilon_a \varepsilon_c F_{c-a} T_c^4$ originate the terms accounting for the heat conduction through the anode body, radiation from the anode front surface, and incident radiation from the cathode respectively. There values are summarized in Appendix III.

Eqs. (12) and (7) represent a system with two unknowns: the surface temperature $T_a$ and the ablation flux $g_{abl}$. An exact solution can be obtained numerically. Alternatively, the arc current $I$ can be obtained directly as a function of the ablation rate by substituting the surface temperature obtained from (8) into (13).

In the next section, we derive analytical solutions. Ablation rate is equal to ablation flux times anode front surface area: $G_{abl} = 4\pi r_a^2 g_{abl}$.

The values of the model parameters are summarized in Appendix III. Note that the model parameters that depend on the arc plasma (the electron temperature, cathode temperature, ion current, and sheath voltage) are almost constant or small compared to the work function term. It allows excluding the arc plasma from the ablation model making it simple.

## V. Arc ablation rate relations for low and high ablation regimes

### V.a. Low ablation regime

In a low ablation regime, for temperatures below $T_{low}$ defined in Eq. (11), the ablation rate term plays a minor role in the heat balance equation (13). Other terms in Eq. (13) are relatively weak functions of the surface temperature $T_a$, hence, Eq. (13) can be linearized in the vicinity of $T_{low}$ to express $T_a$:

$$T_a = V_{eff} I/A + B, \tag{14}$$

where $A = 2.5 C_1 T_{low}^{1.5} r_a^{1.5} + 4 C_2 r_a^2 T_{low}^3 + \pi r_a^2 L \cdot g'(T_{low})$,
$B = T_{low} - \left(C_1 T_{low}^{2.5} r_a^{1.5} + C_2 r_a^2 T_{low}^4 + \pi r_a^2 L \cdot g_0\right)/A$,
$g_0$ is ablation flux at $T_a = T_{low}$ given by Eq. (6), $g'(T_{low})$ is a derivative of ablation flux given by Eq. (9) at $T_a = T_{low}$: $g'(T_{low}) = \frac{g_0}{T_{low}}\left(\frac{Lm}{kT_{low}} - 1\right)$.

For carbon at 500 torr, $g'(T_{low}) \approx 2.2\, g_0/T_{low}$.

$$G'(T_{trans}) \approx 0.7 \cdot G_0 \frac{Lm}{kT_{low}^2}.$$

The ablation rate can be obtained by substituting $T_a$ from (14) into (12):

$$G_{abl} = \pi r_a^2 g_0 \left[\frac{p_0}{p}\exp\left(-\frac{Lm}{k(V_{eff}I/A+B)}\right) + \frac{1}{2}\left(\frac{p_0}{p}\exp\left(-\frac{Lm}{k(V_{eff}I/A+B)}\right)\right)^2\right]. \tag{15}$$

This solution holds for the anode surface temperature $T_a$ below $T_{low}$ and, correspondingly, ablation flux below $g_0$. At $T_a = T_{low}$ and $g_{abl} = g_0$, Eq. (13) transforms to the following relation between the anode radius and the arc current:

$$r_a^2\left[4\pi L g_0 + C_1 T_{low}^{2.5} r_a^{-0.5} + C_2 T_{low}^4 - C_3\right] = V_{eff} \cdot I. \tag{16}$$

A transition to the high ablation regime begins when the arc current becomes higher than one predicted by Eq. (16) (if the anode radius is fixed), or when the anode radius becomes smaller than one predicted by Eq. (16) (if the arc current is fixed).

**V.b. High ablation regime**

In the high ablation regime, the ablation rate grows drastically (several orders of magnitude) while the surface temperature $T_a$ is almost constant, close to the carbon vapor saturation temperature $T_{sat} = -Lm_C/[k\ ln(p/p_0)]$, see Fig. 3. Consequently, the ablation rate in this regime can be determined from substitution $T_{sat}$ in Eq. (13):

$$G_{abl} = (V_{eff} \cdot I + C_3 r_a^2 - C_1 T_{sat}^{2.5} r_a^{1.5} - C_2 r_a^2 T_{sat}^4)/L. \tag{17}$$

Note that the ablation rate in this regime does not depend on the diffusion characteristics, i.e. parameter $g_0$. It is solely determined by the heat balance and the saturation temperature $T_{sat}$ which depends on the background pressure $p$.

## VI. Results and discussion

Multiple experiments have been conducted with carbon arcs for nano-synthesis[4,16,17,18]. In these experiments, arcs were run between two long cylindrical electrodes in an atmosphere of helium at $p = 500\ torr$. Each experimental run was performed at constant arc parameters for a sufficiently long time to consider the arc established so that the parameters don't vary with time. Electrode separation was kept constant by automatically adjusting electrodes positions to compensate for the ablation of a graphite anode ablation and growth of a deposit on a cathode. The ablation rate was accurately measured by weighting the electrodes before and after the experiments. The ablation rate data was collected for various radii of electrodes, their separation distances, and electric currents. This data allows a comprehensive validation of the ablation model.

**VI.a Comparison of analytical formulas for the ablation rate to experimental data from Ref. [18] for variable arc current**

In experiments[18], arc current was varied, whereas other arc parameters were kept unchanged. Anode radius was $r_a = 3mm$. Electrode separation $d$ was maintained in a range 1 mm-2 mm (we use average value $d = 1.5mm$ in the model). Cathodes $4.5mm$ in radius were used, on top of which a deposit of a radius $r_c = 3mm$ formed (see Fig. 7 in Ref. [18] and Fig. 2 in Ref. [1] for a similar arc) corresponding to a view factor between the cathode and the anode $F_{a \to c} = 0.6$.

Experimental data are compared to modeling results in Fig. 6. As evident from the figure, modeling results given by a numerical solution of Eqs. (7) and (13) are in good agreement with the experimental data. A sharp transition between low and high ablation regimes observed in the experiment is well captured by the modeling. Analytical solutions for low and high ablation regimes given by Eqs. (15) and (17) are in a good agreement with both the numerical solution (at corresponding parts of the curve) and the experimental data. The analytical solutions for two regimes cover a substantial part of the curve and generally allow reconstructing the ablation rate behavior without the need for numerical tools.

The analytical solution for low ablation regime given by Eq. (15) was plotted for the surface temperatures $T_a \leq T_{low}$, with $T_{low} = 3760K$ (as defined in Eq. (11)) corresponding to the arc current defined by Eq. (16) determining the right end of the dashed blue curve in Fig. 6. In the low ablation regime, the ablation rate is substantially reduced by the ambient helium; its dependence on $T_a$ and the arc current is exponential, but the values are low. Most of the anode heating is radiated away from its surfaces, the ablation rate is a minor fraction of the anode energy balance.

With the arc current increase, when temperate of the anode front surface $T_a$ approaches the carbon vapor saturation temperature $T_{sat} = 3830K$, then the ablation rate becomes incredibly sensitive to $T_a$ (see Fig. 3). Further increase of the arc current and corresponding increase of the anode heating has very little effect on the front surface temperature. The radiation and heat conduction terms do not change, and all the additional heat goes to the ablation rate increase. The ablation rate is well described by the analytical solution (17) which takes a form of a straight line with a slope $V_{eff}/L$.

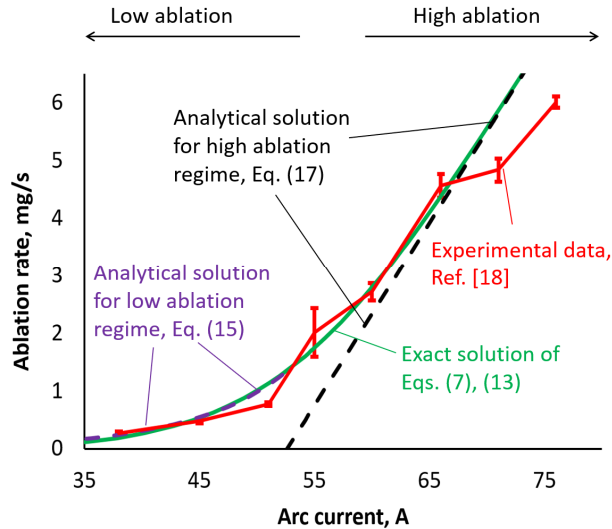

Figure 6. Ablation rate as a function of arc current for fixed inter-electrode gap width d=1.5 mm and anode radius $r_a$=3mm. Good agreement between the numerical solution of Eqs. (7) and (13), analytical solutions for low and high ablation regimes (Eqs. (15) and (17)) and experimental data[18] is obtained.

To study effect of the background pressure *p* on the ablation rate, calculations were performed for various background pressures while other conditions were the same as in the experiments[18]. The results are plotted in Fig. 7 as functions of the arc current. As evident form the figure, straight parts of the curves corresponding to the high ablation regime shift to the right (to higher arc currents) with the background pressure increase. This happens in accordance with Eq. (17). In the high ablation regime, the surface temperature is equal to the saturation temperature $T_{sat}$ which is a logarithmic function of the background pressure. $T_{sat}$ grows with the background pressure thereby increasing radiative and thermal energy losses form the anode and leaving less heat for the ablation. Generally, at any arc current the ablation rate decreases with the background pressure.

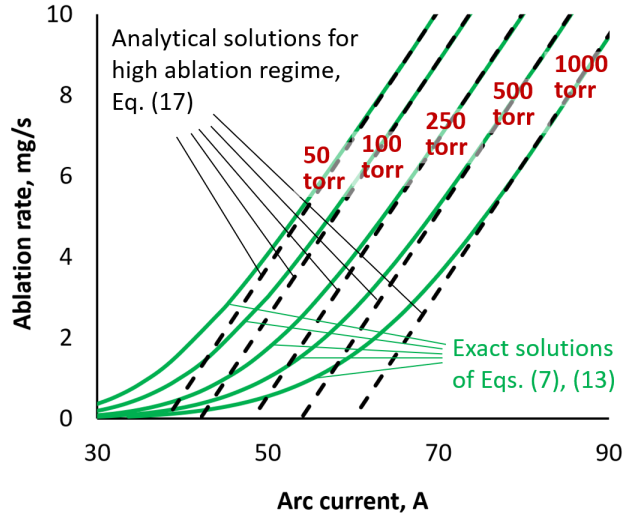

*Figure 7. Ablation rate as a function of arc current for variable background pressure p and fixed inter-electrode gap width d=1.5 mm and anode radius $r_a$=3mm.*

### VI.b. Comparison of the analytical formulas for the ablation rate to experimental data from Refs. [16] and [17] for variable anode radius

In experiments from Refs. [16,17], the anode radius was varied whilst the arc current and inter-electrode gap were constant, 65 A and 1.5mm respectively. The experimental data and modeling results are plotted in Fig. 8. The analytical solution for low ablation regime given by Eq. (15) was plotted for the surface temperatures $T_a \leq T_{low}$ correspond to the anode radii above 3.8 mm as predicted by Eq. (16).

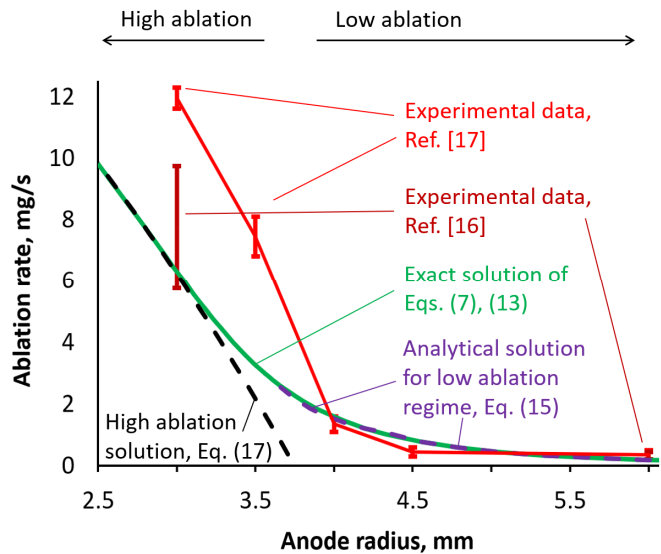

*Figure 8. Ablation rate as a function of the anode radius $r_a$ for arc current I=65 A and inter-electrode gap d=1.5 mm. Good agreement between the numerical solution of Eqs. (7) and (13), analytical solutions for low (15) and high (17) ablation regimes and experimental data is obtained.*

As one can notice, experimental data from different sources slightly varies. This can be explained by measurement errors and some differences in experimental setups. For example, In Ref. [18], relatively narrow cathodes were used, deposit radius was $r_c = 3mm$ corresponding to view factor $F_{a \to c} = 0.6$. In Refs. [16,17], cathodes of radii 6 mm and 25 mm were used having much wider deposit on them (see Fig. 3 in Ref. [17]) corresponding to a view factor of about $F_{a \to c} \approx 1$. Higher view factor implies higher radiation energy flux from the cathode to the anode and higher ablation rate. Unity value was used in the model in this case.

As evident from Fig. 8, the numerical solution of Eqs. (7) and (13) is in a good agreement with the experimental data for a low ablation regime. For high ablation regime, the modeling results lie within the uncertainty range of the experimental data. The abrupt transition between high and low ablation regimes is well captured by the model.

When the anode is wide, energy losses due to radiation from its surface are high and the ablation rate is low. The narrower the anode, the lower are the radiation losses, the higher is the surface temperature $T_a$ and the higher is the ablation rate. When $T_a$ approaches $T_{sat}$, its variation is diminished due to a very high influence on the ablation rate, and the system goes to the high ablation regime described by Eq. (17).

### VI.c. Comparison of the analytical formula for the ablation rate to our experimental data for variable inter-electrode gap

In our arc experiments previously reported in Ref. [25], the inter-electrode gap was varied for the constant arc current of 60 A and anode radius of $r_a = 3mm$. The cathode radius was 4.5 mm corresponding to the deposit radius $r_c = 3mm$. The experimental data is presented in Fig. 8 in comparison with the modeling results and the analytical solution for high ablation regime. Good agreement between the analytical solutions and the experimental data is observed.

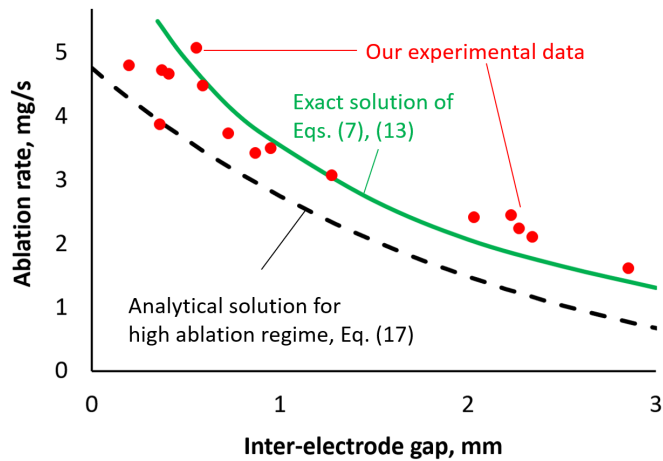

*Figure 9. Ablation rate as a function of the electrode separation distance d for the arc current I=60 A and anode radius $r_a$=3 mm. Good agreement between the numerical solution of Eqs. (7) and (13), analytical solution (17) for high ablation regime and experimental data is obtained.*

As evident from the figure, the inter-electrode gap width noticeably affects the ablation rate: with the gap variation from 0.5 mm to 3 mm the ablation rate changes by a factor of 2.5. The lower is the distance between the electrodes, the higher is the ablation rate. Numerical simulations of carbon arc discharges at similar conditions[24] showed that the near-anode layer in plasma is quite narrow, considerably smaller than 0.5 mm. Therefore, if the gap width is kept above 0.5 mm, its variation should affect neither plasma parameters in the vicinity of the anode nor the heat flux from the plasma to the anode which can be determined as $V_{eff} \cdot I$. The solutions plotted in Fig. 9 show that the arc is operating in a nearly high ablation regime in which the ablation rate does not depend on the carbon diffusion process but solely depends on the anode heat balance. The parameter that affects the anode energy balance and depends on the gap width is the radiation view factor $F_{a \to c}$: $F_{a \to c} = 0.38$ for $d = 3mm$; $F_{a \to c} = 0.85$ for $d = 1mm$. Our modeling shows that the variation of the ablation rate can be explained by the radiation view factor. The narrower the gap, the higher fraction of cathode thermal radiation reaches the anode, and the higher is the ablation rate.

## VII. Summary

A model of the heated graphite surface ablation into a background gas was developed and applied to describe anode ablation in a carbon arc. The deposition of carbon vapor back to the ablating surface was taken into account in the determination of the net ablation flux. A crucial importance of taking a background gas pressure into consideration of the ablation was shown. The manifestation of two distinct ablation regimes observed in arc experiments was explained; analytical expressions for the ablation regimes were derived.

The model was thoroughly validated by comparison to multiple available experimental data on the ablation rate in carbon arcs and a laser ablation system. The model predictions are in good agreement with the experimental data. A rapid transition between high and low ablation regimes is well captured by the model. Analytical expressions are in a good agreement with the model numerical solution.

The ablation model, explanation of the high and low ablation regimes and analytical relations for these regimes are summarized below.

The background gas impedes the flow of the ablated material (i.e. carbon vapor) leading to its accumulation at the ablating surface and resulting in a returning flux of the ablated material back to the surface. An equation for the net ablation flux $g_{abl}$ as a function of the surface temperature $T_a$ was derived,

Eq. (7): $g_{abl} = \frac{p_{sat,C}(T_a)}{v_{th}(T_a)} - \frac{p}{v_{th}(T_a)} \left[1 - exp\left(-\frac{g_{abl}}{g_0}\right)\right]$.

In this equation, the last term accounts for the condensation of the ablated material. Here, $p_{sat,C}(T_a) = p_0 \exp(-Lm_C/kT_a)$ is the carbon saturation pressure, $v_{th}$ is the thermal velocity, $p$ is the background pressure, and $g_0 = m_{C,a} n D_{C-He}/d$ is the diffusion parameter inversely proportional to the condensation distance $d$ which is either the inter-electrode gap width or a distance to the condensation region in case of light beam ablation (see Fig. 2). Values of the model constants are summarized in Appendix III.

This equation was solved numerically for $g_{abl}$ as a function of a surface temperature $T_a$ for typical electric arc parameters (see Fig. 3). In addition to the numerical solution, accurate explicit expression for the surface temperature $T_a$ as a function of $g_{abl}$ was obtained,

Eq. (8): $\frac{1}{T_a} = \frac{1}{T_{sat}} - \frac{k}{Lm_C} \ln\left(\frac{g_{abl}}{p}\sqrt{\frac{2\pi k T_{sat}}{m_C}} + \left[1 - exp\left(-\frac{g_{abl}}{g_0}\right)\right]\right);$

analytical asymptotical solutions for high and low ablation regimes were derived. Here, $T_{sat} = -Lm_C/[k\, ln(p/p_0)]$ is the carbon vapor saturation temperature.

In a low ablation regime, when heating of the ablating surface is low and the surface temperature is substantially below the carbon vapor saturation temperature $T_{sat}$, the ablation rate is strongly reduced by diffusion of the carbon vapor through the background gas. Carbon vapor is almost in equilibrium with the surface; the ablation and deposition fluxes almost cancel out each other resulting in a very small net ablation flux. It was shown that in this regime the ablation flux is proportional to the saturation pressure as would be in case of ablation to vacuum, but with a very small coefficient $\alpha_{diff}$.

Eq. (10): $g_{abl}(T_a) \approx \alpha_{diff}\, p_{sat,C}(T_a)/v_{th}$,

where $\alpha_{diff} = 1.7kT/(p\sigma_{C-He}d)$ is inversely proportional to the background pressure and to the condensation distance $d$. For typical arc parameters, $\alpha_{diff}$ is about 0.003.

This equation was obtained by linearizing the last term in Eq. (7). It is accurate for surface temperatures considerably below $T_{sat}$. More accurate expression was obtained with second order Taylor expansion of the last term in Eq. (7) yielding

Eq. (12): $g_{abl} = g_0 \left(\frac{p_{sat,C}(T_a)}{p} + \frac{1}{2}\left(\frac{p_{sat,C}(T_a)}{p}\right)^2\right).$

This expression is accurate for ablation fluxes up to $g_0$ corresponding to surface temperatures up to $T_{low} = T_{sat}/[1 - 0.46/ln(p/p_0)]$, as defined by Eq. (11), which is slightly below $T_{sat}$.

With the heating increase, when the surface temperature approaches the saturation point of carbon vapor, the carbon pressure at the surface can no longer grow to suppress the ablation. At these conditions, the ablation rate grows drastically with a small variation of the surface temperature. All extra heating applied to the surface virtually does not change its temperature and entirely contributes to the increase of the ablation rate. This is the high ablation regime.

A model of the anode ablation in a carbon arc was developed based on a one-dimensional energy balance in the long and thin electrode. An equation was derived for the temperature of the electrode front surface $T_a$ and the ablation flux $g_{abl}$,

Eq. (13): $L \cdot G_{abl}(T_a) + C_1 T_a^{2.5} r_a^{1.5} + C_2 r_a^2 T_a^4 = V_{eff} \cdot I + C_3 r_a^2.$

Here, $G_{abl} = 4\pi r_a^2 g_{abl}$ is the ablation rate, $r_a$ is the anode radius, and $C_1$, $C_2$ and $C_3$ are constants combined of the material properties and the electrodes view factor. Values of these constants are given in Appendix III. An exact solution for the surface temperature $T_a$ and ablation flux $g_{abl}$ can be obtained numerically. Alternatively, arc current $I$ can be obtained directly as a function of ablation rate by substituting surface temperature obtained from (8) into (13).

Analytical expressions for the ablation rate in low and high ablation regimes were derived. In the low ablation regime, Eq. (13) can be linearized in the vicinity of $T_{low}$ to obtain $T_a$. This $T_a$ can be substituted in Eq. (12) yielding following relation for the ablation rate,

Eq. (15): $G_{abl} = \pi r_a^2 g_0 \left[ \frac{p_0}{p} \exp\left(-\frac{Lm}{k(V_{eff}I/A+B)}\right) + \frac{1}{2}\left(\frac{p_0}{p}\exp\left(-\frac{Lm}{k(V_{eff}I/A+B)}\right)\right)^2 \right]$.

Here, $A = 2.5 C_1 T_{low}^{1.5} r_a^{1.5} + 4 C_2 r_a^2 T_{low}^3 + 2.2\, \pi r_a^2 L \cdot g_0 / T_{low}$,
and $B = T_{low} - \left(C_1 T_{low}^{2.5} r_a^{1.5} + C_2 r_a^2 T_{low}^4 + \pi r_a^2 L \cdot g_0\right)/A$.

The arc operates in a low ablation regime until the anode surface temperature $T_a$ below $T_{low}$ and, correspondingly, ablation flux is below $g_0$. At these parameters, the anode radius $r_a$ and the arc current $I$ are tied by a following relation to satisfy Eq. (12):

Eq. (16): $r_a^2 \left[ 4\pi L g_0 + C_1 T_{low}^{2.5} r_a^{-0.5} + C_2 T_{low}^4 - C_3 \right] = V_{eff} \cdot I$.

Here, $C_1 = \pi\sqrt{4/5 \cdot \sigma \varepsilon_a \lambda_a}$, $C_2 = \pi\sigma\varepsilon_a(1-\alpha_r)$ and $C_3 = \pi\sigma\varepsilon_a\varepsilon_c F_{c-a} T_c^4$ are constants combined of the material properties and the radiation view-factor. Their values are summarized in appendix III.

A transition to the high ablation regime begins when the arc current becomes higher than one predicted by Eq. (16) (if the anode radius is fixed), or when the anode radius becomes smaller than one predicted by Eq. (16) (if the arc current is fixed).

In the high ablation regime, surface temperature $T_a$ is equal to the saturation temperature $T_{sat}$. Variation of the anode heating virtually does not affect $T_a$ and entirely contributes to the change of the ablation rate, yielding

Eq. (17): $G_{abl} = \left(V_{eff} \cdot I + C_3 r_a^2 - C_1 T_{sat}^{2.5} r_a^{1.5} - C_2 r_a^2 T_{sat}^4\right)/L$.

Effect of the background pressure $p$ on the ablation rate was studied. The results show that at higher background pressure a transition to the high ablation regime requires higher arc current. In the high ablation regime, the surface temperature is equal to the saturation temperature $T_{sat}$ which grows logarithmically with the background pressure thereby increasing radiative and thermal energy losses form the anode and leaving less heat for the ablation. Generally, at any arc current the ablation rate is the higher the lower is the background pressure.

## Acknowledgements

The authors would like to thank Dr. Valerian Nemchinsky (Keiser University, FL), Dr. Yevgeny Raitses (PPPL) and Jian Chen (PPPL) for fruitful discussions. The theoretical work was supported by the US Department of Energy (DOE), Office of Science, Fusion Energy Sciences. The arc experiments were supported by the US DOE, Office of Science, Basic Energy Sciences, Materials Sciences and Engineering Division.

## Data Availability Statement

The data that support the findings of this study are available from the corresponding author upon a reasonable request.

## Appendix I. Parameters for the graphite ablation flux

The ablation flux relations (1) and (3) are formulated for single gas species. However, several carbon species ablate from a surface of graphite: atoms C and molecules $C_2$ and $C_3$, as predicted by thermodynamic modeling[1,42,43] and observed in experiments[1,42,44]. A detailed model of the ablation process would require separate consideration of species C, $C_2$, and $C_3$ (as done in Ref. [21]) with individual ablation rates. Luckily, the value of the latent heat per one particle ($L \cdot m$) in the exponent of saturation pressure expression (2) is close for C, $C_2$, and $C_3$ (see Refs. [21,44,45,46,47]), and mass of ablating particles is under a radical in thermal velocity expression. This allows a rather accurate determination of total ablation flux with a single relation (3). Following values of the material parameters were used in the present model: $L \cdot m_C = 1.2 \cdot 10^{-18} \, J$, $p_0 = 4.8 \cdot 10^{14} \, Pa$ (corresponding to $p_{satur,C}$ equal to 1 atm. at 3900 K), $m_C = 4.0 \cdot 10^{-26} kg$ (corresponding to the mass of a $C_2$ molecule).

## Appendix II. Diffusion equation for the transport of carbon gas

Even though not only carbon atom atoms but also carbon molecules $C_2$ and $C_3$ ablate from a graphite surface (as discussed in Appendix I), the molecules are only present in very thin regions near the arc electrodes where the temperature is low[42] and, therefore, don't play important role in the carbon transport in the arc volume (from the anode towards the cathode). The same can be said for the carbon transport in a laser ablation system[7,8], where carbon molecules $C_2$ and $C_3$ were only detected at a distance of about 1mm from the ablating surface where carbon condenses and nanoparticles form. Thereby, diffusion of carbon atoms only is considered here.

Diffusion of carbon atoms through other components of a carbon arc or a light beam ablation plasma is accurately described by the Stefan-Maxwell equation[26,27,28,29] derived from the kinetic theory of gases. In a one-dimensional form this equation reads:

$$\sum_j \frac{n_C n_j kT C_{C-j}}{n D_{C-j}} (v_C - v_j) = -\frac{dp_C}{dx} + Y_C \frac{dp}{dx} - C_C^{(e)} n_C k \frac{dT}{dx}. \tag{AII.1}$$

Here, index $j$ denotes a gas mixture component other than carbon atoms (helium atoms (He), carbon ions (i) and electrons (e)), $x$ is a coordinate in a direction perpendicular to the ablating surface, $p$ is the mixture pressure, $T$ is the temperature (assumed equal for all species), $p_C = n_C kT$ is the partial pressure of carbon gas, $k$ is the Boltzmann constant, $Y_C$ is the mass fraction of carbon atoms, $v_i$ are individual mass-averaged velocities of different species, $n_i$ are species densities, $C_{C-j}$ and $C_C^{(e)}$ are kinetic coefficients, and $D_{C-j}$ are binary diffusion coefficients given by

$$n D_{C-i} = \frac{3\pi}{32} \sqrt{\frac{8kT}{\pi m_{C-j}}} \frac{1}{\sigma_{C-j}}, \quad m_{C-j} = \frac{m_{C,a} m_j}{m_{C,a} + m_j}, \tag{AII.2}$$

where $\sigma_{C-j}$ is a collision cross-section of carbon atoms with species $j$; $m_{C,a}$ is the mass of carbon atoms; $m_j$ are other species masses, and $n = \sum n_i = p/kT$ is the gas mixture density.

In general, Eq. (AII.1) describes momentum balance for carbon atoms: friction of carbon atoms with other mixture species due to their relative motion is compensated by the gradients of the species pressures and temperatures. Eq. (AII.1) has a complex form, however, only two of its terms are large, the rest can be neglected as discussed blow.

Kinetic coefficient $C_C^{(e)}$ is of order of the ionization degree[27,29], which is typically below 0.1 in a carbon arc[24], hence, the last term in Eq. (AII.1) is small in comparison to the first term in the right-hand side (RHS) of Eq. (AII.1) and can be neglected.

Variation of the carbon atoms partial pressure $p_C$ is by many orders of magnitude higher than variation of the mixture pressure $p$. $p_C$ varies from its maximum at the ablating surface where it is comparable to $p$ to almost zero at region where carbon condensation happens (a cathode surface in a carbon arc or a volumetric condensation region in case of a light beam ablation, see Fig. 2). From the Bernoulli equation, variation of the mixture pressure $p$ is of order of $\rho v^2$ or $g^2/\rho$, where $\rho$ is the mixture density ($\approx 0.02 \ kg/m^3$ for carbon gas at 500 torr and 5 000 K typical for carbon arc[24]), $v$ is the mass-averaged flow velocity and $g$ is the ablation flux (below $0.3 \ kg/m^2s$ corresponding to 10 mg/s from a 3mm radius surface, as in the arc and light-beam ablation experiments). With these parameters, variation of the mixture pressure $p$ is about 5 Pa, much lower than $p$ itself (67 kPa) and the variation of $p_C$. With this said, second term in the RHS of Eq. (AII.1) can be neglected.

For convenience, left-hand side (LHS) of Eq. (AII.1) can be reformulated in terms of particle fluxes:

$$\sum_j \frac{C_{C-j}}{nD_{C-j}}(p_j J_C - p_C J_j) = -\frac{dp_C}{dx} \tag{AII.3}$$

Here, $J_j = n_j v_j = g_j/m_j$ is the particle flux of species $j$, $p_j = n_j kT$ is the species partial pressure. With three mixture species $j$ other than carbon atoms (helium atoms (He), carbon ions (i) and electrons (e)), there are six terms in the LHS of Eq. (AII.2). However, among these terms only the one with $p_{He} J_C$ is important; other terms are small and can be neglected, as is shown below.

Collision cross sections between the species are: $\sigma_{C-He} \approx 3 \cdot 10^{-19} \ m^2$ [48], $\sigma_{C-e} \approx 2 \cdot 10^{-19} \ m^2$ [49,50] (for low-energy elastic collisions), $\sigma_{C-i} \approx 7 \cdot 10^{-19} \ m^2$ [51,52] (charge exchange cross section for the carbon atom-ion collisions). The binary masses are: $m_{C-He} = 1.5 \cdot 10^{-26} kg$, $m_{C-e} = 9 \cdot 10^{-31} kg$, $m_{C-i} = 10^{-26} kg$. Binary diffusion coefficients between carbon atoms and other species, as defined by (AII.2), are weak functions of temperature. For typical temperature in electric arc, 5 000 K, the diffusion coefficients are equal to: $nD_{C-He} = 5 \cdot 10^{21} \ m^2/s$, $nD_{C-i} = 2 \cdot 10^{21} \ m^2/s$, $nD_{C-e} = 10^{24} \ m^2/s$.

For a typical carbon arc with a current of 70 A, ablation rate 3 mg/s, and anode radius 3 mm, particle fluxes of carbon atoms and electrons, correspondingly, are $J_C = 5 \cdot 10^{24} \ m^{-2}$, $J_e = 1.5 \cdot 10^{25} \ m^{-2}$. Ion flux can be estimated as $J_i \approx J_e \sqrt{m_e/m_C} = 10^{23} \ m^{-2}$. In light-beam ablation systems, ion and electron fluxes are negligible. The background gas, i.e. helium, is not moving, hence $J_{He} = 0$.

Partial pressures of carbon $p_C$ and helium $p_{He}$ are of the same order, partial pressures of ions $p_i$ and electrons $p_e$ are by more than an order of magnitude lower, corresponding to the ionization degree below 0.1 [24].

Kinetic coefficients $C_{C-j}$ are of order of unity[27,29]; $C_{C-He} = 1$.

With the above-written estimates, ratios of the terms in the LHS of (AII.3) can be determined:

$$\frac{p_{He}J_C}{nD_{C-He}} \Big/ \frac{p_C J_{He}}{nD_{C-He}} \Big/ \frac{p_i J_C}{nD_{C-i}} \Big/ \frac{p_C J_i}{nD_{C-i}} \Big/ \frac{p_e J_C}{nD_{C-e}} \Big/ \frac{p_C J_e}{nD_{C-e}} \approx$$

$$\approx 1 \quad / \quad 0 \quad / \quad 0.01 \quad / \quad 0.03 \quad / \quad 3 \cdot 10^{-5} \quad / \quad 0.01 \,. \tag{AII.4}$$

As evident from relation (AII.4), all terms in the LHS of (AII.3) except for the term containing $p_{He}J_C$ are small and can be neglected, thereby reducing (AII.3) to

$$\frac{p_{He}}{nD_{C-He}} J_C = -\frac{dp_C}{dx}, \tag{AII.5}$$

where $J_C = g_C/m_{C,a}$. In a one-dimensional consideration, carbon flux $g_C$ is constant (does not vary with $x$) equal to the ablation flux $g_{abl}$. According to Eq. (AII.5), the major mechanism for the carbon gas motion is its diffusion through stationary helium driven by the carbon gas pressure gradient.

Since the variation of the gas mixture pressure $p$ is small and the fraction of electrons and ions in the gas mixture is low, the sum of carbon and helium partial pressures can be taken constant, equal to the background pressure:

$$p_C + p_{He} \approx p = const. \tag{AII.6}$$

Using (AII.6), $p_{He}$ can be excluded from (AII.5):

$$\frac{p - p_C}{nD_{C-He}} \frac{g_{abl}}{m_{C,a}} = -\frac{dp_C}{dx} \tag{AII.7}$$

Equation (AII.7) can be solved analytically for $p_C$:

$$p_C = p - C \exp\left(-\frac{x}{nD_{C-He}} \frac{g_{abl}}{m_{C,a}}\right), \tag{AII.8}$$

where $x$ is the distance from the ablation surface, and $C$ is a constant determined from the boundary conditions. Zero carbon density can be used as a boundary condition at a location $x = d$ corresponding to either a cathode surface in an electric arc or a condensation layer in the laser/solar ablation case, see Fig. 2. With this, the constant $C$ can be determined yielding following carbon density profile:

$$p_C = p - P \exp\left(\frac{d-x}{nD_{C-He}} \frac{g_{abl}}{m_{C,a}}\right), \tag{AII.9}$$

Substituting $x = 0$ into (AII.9) yields carbon density at the ablation surface, $p_{C,a}$:

$$p_{C,a} = p\left[1 - \exp\left(-\frac{g_{abl}}{g_0}\right)\right], \tag{AII.10}$$

where $g_0 = \frac{m_{C,a} n D_{C-He}}{d}$.

# Appendix III. Values of the model parameters

| Parameter | Value | Reference |
|---|---|---|
| $L$ | 3·10 | [44], [45], [46], [47] |
| $L \cdot m_C$ | 1.2·10 | |
| $p_0$ | 4.8·10 | |
| $\varepsilon_a = \varepsilon_c$ | 0.8* | [17], [53] |
| $T_c$ | 3400 K** | [29] |
| $\lambda_a$ | 14 W/m/K*** | [17], [54] |
| $V_{eff}$ | 9.5 V | |
| $m_C$ | 4.0·10$^{-26}$ | Mass of a $C_2$ molecule |
| $m_{C,a}$ | 2.0·10$^{-26}$ | Mass of a carbon atom |
| $m_{C-He}$ | 5.0·10$^{-27}$ | Binary mass of a carbon and helium atoms |
| $\sigma_{C-He}$ | $3 \cdot 10^{-19}\ m^2$ | [48] |
| $nD_{C-He}$ | $5 \cdot 10^{21}\ m^{-1}s^{-1}$ | [48] |
| $F_{c-a}$ | = 0.6 for a cathode ($r_c$=4.5mm) and $d = 1.5mm$ as in Ref. [18]; see Ref. [33]. = 1 for a wide cathode as in Refs. [16] and [17] | |
| $\alpha_r$ | = 0.06 for a narrow cathode, as in [33]; = 0.16 for a wide cathode, as in [16] and [17] | |
| $C_1$ | 0.00225 W K$^{-2.5}$m$^{-1.5}$ | |
| $C_2$ | = 1.35·10$^{-7}$ W K$^{-4}$m$^{-2}$ for a narrow cathode, as in [33]; = 1.2·10$^{-7}$ W K$^{-4}$m$^{-2}$ for a wide cathode, as in [16], [17] | |
| $C_3$ | = 9.3·10$^6$ W m$^{-2}$ for a narrow cathode, as in [33]; = 1.5·10$^7$ W m$^{-2}$ for a wide cathode, as in [16], [17] | |
| $T_{low}$ | $3750K$ at 500 torr; $3630K$ at 230 torr | |
| $T_{sat}$ | $3830K$ at 500 torr; $3700K$ at 230 torr | |

*This value is for both anode graphite and cathode deposit[17,53].

**Cathode temperature is very weakly (logarithmically) dependent on the arc current[24,29]; constant value of 3400 K is used in the model.

***The value for graphite at high temperature is taken, corresponding to a region near the front surface where a major part of heat transfer takes place.